# Coexistence of Intrinsic Superconductivity and Topological Insulator State in Monoclinic Phase WS$_2$

Yuqiang Fang[†,1,4,10], Jie Pan[†,1], Dongqin Zhang[†,2,3], Dong Wang[1], Hishiro T. Hirose[7], Taichi Terashima[7], Shinya Uji[7], Yonghao Yuan[5,6], Wei Li[5,6], Zhen Tian[8], Jiamin Xue[8], Yonghui Ma[9], Wei Zhao[1], Qikun Xue[5], Gang Mu[*,9], Haijun Zhang[*,2,3], Fuqiang Huang[*,1,4]

[1]State Key Laboratory of High Performance Ceramics and Superfine Microstructure, Shanghai Institute of Ceramics, Chinese Academy of Science, Shanghai (P. R. China). [2]National Laboratory of Solid State Microstructures and School of Physics, Nanjing University, Nanjing (P. R. China). [3]Collaborative Innovation Center of Advanced Microstructures, Nanjing, (P. R. China). [4]State Key Laboratory of Rare Earth Materials Chemistry and Applications, College of Chemistry and Molecular Engineering Peking University, Beijing (P. R. China). [5]State Key Laboratory of Low-Dimensional Quantum Physics, Department of Physics, Tsinghua University, Beijing, (P. R. China). [6]Collaborative Innovation Center of Quantum Matter, Beijing, (P. R. China). [7]National Institute for Materials Science, Tsukuba, Ibaraki (Japan). [8]School of Physical Science and Technology, ShanghaiTech University, Shanghai, (P. R. China). [9]State Key Laboratory of Functional Materials for Informatics, Shanghai Institute of Microsystem and Information Technology, Chinese Academy of Sciences, Shanghai, (P. R. China). [10]University of Chinese Academy of Sciences, Beijing (P. R. China). [†]These authors contributed equally to this work.

*e-mail: mugang@mail.sim.ac.cn; zhanghj@nju.edu.cn; huangfq@mail.sic.ac.cn.

## Abstract

**Recently, intriguing phenomena of superconductivity, type-II Weyl semimetal or quantum spin Hall states were discovered in metastable 1T'-type VIB-group transition metal dichalcogenides (TMDs). Here, we report that monoclinic phase WS$_2$ was discovered and synthetized in our experiments. The intrinsic superconducting transition was observed in monoclinic WS$_2$ with a transition temperature $T_c$ of 8.8 K which is the highest among previously reported TMDs without any fine-tuning process. Interestingly, topological insulator state, defined by topological invariant $Z_2$, was also discovered with a single Dirac cone on the surface, which is different from all topological states reported in TMDs. Further, the electronic structure was found to have a strong anisotropy by Shubnikov–de Haas oscillations and first-principles calculations. Our findings reveal that monoclinic WS$_2$ might be a new topological superconductivity candidate with a strong anisotropy.**

## Introduction

Recently, VIB-group layered transition metal dichalcogenides (TMDs), $MX_2$ ($M$= Mo, W; $X$= S, Se, Te), attracted extensive attention due to rich physiochemical properties, ranging from catalysis (*1-3*), topological states (*4-16*), valley polarization (*17-22*) even to superconductivity (*23-28*). These multiple electronic properties essentially originate from varied crystal structures of TMD materials. The typical crystal structure in TMD materials is the 2H-type structure with [*X-M-X*] atoms in ABA stacking in each monolayer (**Figure 1a**). Usually, 2H $MX_2$ materials are semiconducting, such as, 2H $MoS_2$, where the valley polarization was widely studied (*17-19*). Also, ising superconductivity was observed when the TMD materials are reduced down to a few or even one layer (*29, 30*). Another typical structure is 1T-type structure with [X-M-X] atoms in ABC stacking in each monolayer (**Figure 1b**). The 1T structure is known to undergo a spontaneous lattice distortion to the 1T' structure where the M atoms form characteristic in-plane M-M zigzag chains in each monolayer (**Figure 1f**) (*31, 32*) compared with no *M-M* bonds in the 1T structure (**Figure 1c**). The two stacking modes of 1T' monolayers were found, which were named as 1T' phase (**Figure 1d**) and $T_d$ (**Figure 1e**) phase, respectively. The inversion symmetry is broken in the $T_d$ structure. The type-II Weyl semimetal state (*10, 11*) was predicted in $T_d$ $MoTe_2$ and $WTe_2$, accompanied by a large non-saturating magnetoresistance (*33*) or superconductivity (*34, 35*). In addition, due to heavy transition metal *M* (Mo and W) atoms, the spin-orbit coupling (SOC) effect plays a significant role in the topological nature of TMDs. The quantum spin Hall (QSH) effect was predicted in the monolayer 1T'-type $MX_2$ (*4*), and some experiments provided strong evidences for the QSH states in the monolayer 1T' $WTe_2$ (*5-7, 9*) and 1T' $WSe_2$ (*8*).

Here we report a new monoclinic phase for $WS_2$, labelled as the 2M phase, which is firstly reported as far as we know. The monolayer of 2M $WS_2$ is the 1T' structure, but its crystal structure has a different packing manner of 1T' monolayers, compared with the $T_d$ or 1T' structure, shown in **Figure 2a**. The 2M $WS_2$ displays the intrinsically highest superconducting transition temperature $T_c$ of 8.8 K among all the reported TMD materials without any fine-tuning process, such as, external pressures or gating processes. Interestingly, our first-principles calculations found that the normal state of 2M WS$_2$ is a topological insulator state, defined by topological invariant Z$_2$, which is completely different from Weyl semimetal state previously discovered in T$_d$ MoTe$_2$ and T$_d$ WTe$_2$. The coexistence of the superconductivity and the topological insulator state makes 2M WS$_2$ be a promising candidate for the topological superconductor.

## Results and Discussion

### Crystal structure characterizations

In our experiment, the 2M WS$_2$ single crystals were prepared by deintercalation of potassium cations from K$_{0.7}$WS$_2$ crystals (*36*), where the K content was determined by the EDS analysis (**Table S1**). The corresponding experimental details were illustrated in the **Supplemental Information**. The 2M WS$_2$ single crystal crystallizes in the monoclinic space group C$_{2/m}$ in which the inversion symmetry is preserved. The lattice parameters of 2M WS$_2$ are *a*= 12.87 Å, *b*= 3.23 Å, *c*= 5.71 Å, *α*= *γ*= 90°, *β*= 112.9°, respectively. The S-W-S sandwich monolayer has the 1T' structure with distorted [WS$_6$]$^{8-}$ octahedrons sharing edges along *bc* plane and the W-W zigzag chains in the *b* direction. The 2M WS$_2$ crystals grew mainly along the *c* direction to form a thin narrow ribbon (inset of **Figure 2b**). The structure parameters from the single crystal X-ray diffraction of the 2M WS$_2$ are summarized in **Table S2-S5**. The SEM image (**Figure S1a**) shows the lateral size of as-prepared 2M WS$_2$ crystals reaches several hundred micrometers. Moreover, the element analysis by the energy dispersion spectrum (EDS) measurement (**Figure S1b**) proves no residual of potassium element in obtained 2M WS$_2$ crystals after sufficient oxidation treatment. The measured powder X-ray diffraction pattern (**Figure 2b**) of the sample matches well with the simulated one, further indicating the high purity of the samples.

Although sharing the identical 1T' monolayer with 1T' or T$_d$ WTe$_2$ (*37*), 2M WS$_2$ adopts a different stacking mode. For example, the Te-W-Te monolayers of 1T' or T$_d$ WTe$_2$ stack along the *c* axis through the 2$_1$ glide axis (**Figure 1d** and **1e**), while the S-W-S monolayers of 2M WS$_2$ stack along the *a* axis through the translation operation, shown in **Figure 2a**. Naturally, 2M WS$_2$ expects to host different electronic structure and physical properties from 2H, 1T, 1T' or T$_d$ TMD materials due to the different crystal structure.

The atomically resolved scanning tunneling microscopy (STM) topography clearly reveals the characteristic W-W zigzag chains in the 2M WS$_2$ crystal (**Figure 2d**). Moreover, the image and the FFT pattern (inset of **Figure 2d**) exhibit the rectangular lattice with lattice parameters of *b*= 3.29 Å, *c*= 5.66 Å respectively, which are consistent with those obtained from the single-crystal X-ray diffraction. The Raman spectrum (**Figure 2c**) of the 2M WS$_2$ crystals shows more Raman peaks at the low frequency region, demonstrating the low-symmetry characteristic of 2M WS$_2$ (*38*). The absence of Raman peaks located at 350 cm$^{-1}$ and 417 cm$^{-1}$ confirms that there is no 2H phase in the as-prepared sample.

**Physical transport measurements**

Subsequently the superconductivity and physical parameters of 2M WS$_2$ crystals were characterized through fundamental physical transport measurements. **Figure 3a** shows the temperature-dependent magnetic susceptibility for the 2M WS$_2$ crystals in the zero-field cooling (ZFC) and field cooling (FC) measurements under the magnetic field of 5 Oe. The diamagnetic transition appears at 8.8 K in both the ZFC and FC curves. Heat capacity measurement also shows the abrupt increase near the 8.8 K (**Figure S2**), which confirms the bulk superconductivity in the 2M WS$_2$ crystals. Furthermore, the magnetic hysteresis loop at 2 K demonstrates 2M WS$_2$ crystals as a typical type-II superconductor. The $\rho_{xx}$-T curve of 2M WS$_2$ single sheet under zero magnetic field shows a metallic behavior in the normal state (**Figure 3b**). The rather large residual resistivity ratio RRR [$\rho_{xx}$ (300K)/ $\rho_{xx}$ (0K)]= 227.5 indicates the high quality of our sample. The superconducting transition started at 8.9 K, and reached zero resistivity at 8.7 K, being consistent with the result in the magnetic measurement. Notably the $T_c$ of 2M WS$_2$ crystals is the highest among the intrinsic TMD materials without external pressures and gating processes.

**Figure 3c** and **3d** illustrate the temperature dependence of resistivity under different magnetic fields along the out-of-plane and in-plane directions. The superconductivity at 2 K is completely suppressed under out-of-plane magnetic field of 3 T while the superconductivity still exists under in-plane magnetic field of 9 T. The anisotropy ratio of the upper critical magnetic field $H_{c2}$ is $\gamma = B_{c2}^{bc}/B_{c2}^{a*} = 11.6 \sim 13.7$ in the $B_{c2}$-$T_c$ curve (**Figure S3**) where $T_c$ is defined by the criterion of 98% of normal-state resistance. This value indicates a large anisotropy of superconductivity for 2M WS$_2$ crystals. Furthermore, the temperature dependence of the out-of-plane critical field $B_{a* c2}$ can be well fitted by the Ginzburg-Landau (GL) theory $B_{a* c2} = [\Phi_0/(2\pi\xi_{bc}^2)](1-T/T_c)$, with $\Phi_0 = h/2e$, being the superconducting flux quantum (*39*). The in-plane GL superconducting coherence length is thus estimated as $\xi_{bc} \sim 8.8$ nm (at T = 0).

The Hall effect measurement (**Figure 3e**) shows that the hole-type carrier is dominant in the 2M WS$_2$ crystals. However, the nonlinear behavior of the $\rho_{xy}$-B curve in the low-temperature region indicates more than one type of charge carrier coexist in this system (*40*). For simplicity, the estimation of carrier concentration is based on the single-carrier Drude band model. The density of carrier $n_{e,h}$ is estimated to be $8.58\times10^{20}$ cm$^{-3}$ at T= 10 K, and shows positive correlation with temperature on the whole, exhibiting a semimetallic behavior. The $n_{e,h}$ is two order of magnitudes larger than other semimetals reported such as Cd$_3$As$_2$ (*41*) ($\sim 4.4\times10^{18}$ cm$^{-3}$ at 5 K) and NbP (*40*) ($\sim 1.5\times10^{18}$ cm$^{-3}$ at 1.5 K). Consequently, the carrier mobility $\mu$ calculated from the equation $\sigma=ne\mu$ (where $e$ is the elementary charge) is relatively small (the highest value of $\mu$ is 6745.4 cm$^2\cdot$V$^{-1}\cdot$s$^{-1}$ at 10 K). Due to the effect of the lattice scattering, the $\mu$ decreases rapidly with the increase of temperature.

**First-principles calculations and Shubnikov–de Haas (SdH) oscillations**

We performed first-principles calculations for the electron-phonon coupling of the 2M WS$_2$. The calculated phonon dispersion has no negative frequency indicating the dynamic stability of crystal structure of the 2M WS$_2$ (**Figure S4**). The large frequency gap 5.3 Thz and 6.3 Thz at Gamma point originates from the large mass of W atom (*42*). **Figure S5** shows the calculated phonon density of states, $G(\omega)$, and electron-phonon spectral function, $\alpha^2F(\omega)$. The superconducting transition temperature $T_c$ can be estimated by the Allen and Dynes modified McMillan formula as following:

$$T_c = \frac{\langle \omega \rangle}{1.20} \exp(-\frac{1.04(1+\lambda)}{\lambda - \mu^*(1+0.62\lambda)})$$

The Coulomb parameter μ* is defined as a common value of 0.1 (*43*). The $T_c$ is estimated to be about 2 K, being one fourth of the experimental value. This discrepancy was also found in Fe-based superconductors (*43*). Therefore, the bulk 2M $WS_2$ seems to be a non-conventional electron-phonon superconductor.

The electronic structure analysis of 2M $WS_2$ was performed from both first-principles calculations and SdH oscillations. The band structures with and without the SOC effect are shown in **Figure 4a** and **4b**, which present a semi-metallic picture without a full band gap at the Fermi level. There are hole-like bands around Y point and electron-like bands around Γ point in the band structure. The corresponding Fermi surfaces in the Brillouin zone (BZ) are also calculated, seen in **Figure 4c**, showing the coexistence of hole and electron pockets. The multi-band electronic structure is consistent with the Hall transport measurements. The density of states (DOS) consist of the semimetallic band structures at the Fermi level (**Figure S6**).

**Figures 4d** and **4e** show SdH oscillations and their Fourier transforms, respectively. Both samples #1 and #2 are measured under a magnetic field along $a^*$-axis but with different direction of electrical current (**Figure 4f**). Both samples show consistent frequencies of 80 T ($F_†$) and 807 T. The frequency around 807 T is possibly attributed to the $\gamma_1$ branch, which corresponds to the electron pocket. However, the possibility of δ branch cannot be excluded because of the predominant hole conduction. Moreover, there are two additional frequencies, 150 T and 167 T, which are intense in sample #2 but are absent in sample #1 (**Figure 4e**). These two frequencies are attributed to $\alpha_1$ and $\alpha_2$ branches of the flattened-torus-shaped part of the hole Fermi surface (**Figure 4c**). The flattening along *b*-axis leads to a significantly anisotropic contribution to conductivity: huge (tiny) along *b*- (*c*-) axis. Thus, the difference in the SdH intensities between samples #1 and #2 can be attributed to the difference in the current direction: the electrical current is nearly parallel to the *b*-axis in sample #2. The validity of the assignments of $\alpha_1$ and $\alpha_2$ can be also confirmed by comparing the angular dependence of frequencies between the experiments and the simulations shown in **Figure 4g**. Based on the above discussion, the observed SdH oscillations, except $F_†$ (see **Methods**), are in good agreement with the calculated electronic structure.

**Topological insulator state**

Though the 2M WS$_2$ is found to be a semimetal, it is still possible to define the topology for its electronic structure through a curving Fermi level schematically shown by the green dotted line in **Figure 4b**, because there is a direct band gap between valence and conduction bands at each k point. For example, 36 valence bands are below the curving Fermi level. Therefore, the topology can be calculated for these 36 valence bands in the BZ. Because both the inversion and time-reversal symmetries are preserved in 2M WS$_2$, the parity can be used to calculate the topological invariant $Z_2$. The parity products of the 36 valence bands at each time-reversal-invariant point are listed in **Table 1**. We see that the parity product is odd at Γ point and even at all other time-reversal-invariant points, leading to a nontrivial topology with $Z_2$ (1;000). At the Γ point, there is a band inversion between the *p* orbital of S atoms and the *d* orbital of W atoms, seen in **Figure 4a** and **b**, which is the key for the nontrivial topology.

The topological surface states have to appear on the boundary due to the topologically nontrivial bulk band structure. The surface states on the cleavage (100) surface are calculated by the Green's functions method combing with maximally localized Wannier functions (*44, 45*), shown in **Figure 5a**. There is a single Dirac cone at the Γ point for the surface states. Because the semimetallic bulk electronic structure, the topological surface states partially mixed with bulk states. In order to clearly present the surface and bulk states, the schematics are given in **Figure 5b**. Around the Dirac point, we calculate the Fermi surfaces and the spin texture at four different energy levels (E1, E2, E3 and E4) marked in **Figure 5b**. The Fermi surfaces indicate that both bulk and surface states have a strong anisotropy in the (100) plane. When the Fermi level is close to the Dirac point, the spin texture goes along the Fermi surface of surface states contributing a Berry phase π, seen in **Figure 5d-f**. The surface states expect to be superconducting due to the natural proximity effect between bulk and surface states. Therefore, 2M WS$_2$ becomes a promising topological superconductivity candidate because of the coexistence of the superconductivity and topological surface states (*46,*

*47*).

## Conclusions

In summary, we firstly synthesized the 2M WS$_2$ compound with the $T_c$ of 8.8 K through the oxidation process of the precursor K$_x$WS$_2$ crystals. The 2M WS$_2$ adopts a new crystal structure in TMD materials, different from previously reported 1T' and T$_d$ structures. Physical transport measurements and first-principles calculations demonstrate the coexistence of the superconductivity and the topological surface state, which makes the 2M WS$_2$ be a new promising topological superconductor candidate with a strong anisotropy. Therefore, our findings provide an important material platform on TMD materials for the further study of topological superconductivity.

## Materials and Methods

**Sample Preparation.**

2M WS$_2$ single crystals were prepared by deintercalation of interlayer potassium cations from K$_{0.7}$WS$_2$ crystals. For the synthesis of K$_{0.7}$WS$_2$, K$_2$S$_2$ (prepared via liquid ammonia), W (99.9%, Alfa Aesar) and S (99.99%, Alfa Aesar) were mixed by the stoichiometric ratios and ground in an argon-filled glovebox. The mixtures were pressed into a pellet and sealed in the evacuated quartz tube. The tube was heated at 850 ºC for 2000 min and slowly cooled to 550 ºC at a rate of 0.1 ºC/min.

The synthesized K$_{0.7}$WS$_2$ was oxidized chemically by an excess of 0.01 mol/L K$_2$Cr$_2$O$_7$ in aqueous H$_2$SO$_4$ at room temperature for 1h. Finally, the 2M WS$_2$ crystals were obtained after washing in distilled water for several times and drying in the vacuum oven at room temperature.

**Crystal structure determination.**

Suitable crystals were chosen to perform the data collections. Single-crystal X-ray diffraction was performed on a Bruker D8 QUEST diffractometer equipped with Mo *K*α radiation. The diffraction data were collected at room temperature by the ω- and φ-scan methods. The crystal structure was solved and refined using APEX3 program. Absorption corrections were performed using the multiscan method (SADABS).The detailed crystal data and structure refinement parameters are summarized in **Table S2-5**.

**Raman measurement.**

Raman spectra were performed under the laser of 633nm in the Renishaw inVia Raman Microscope, and the parameter of ND filter is 10%.

**Physical transport measurement.**

Magnetic measurements was taken in the Physical Properties Measurement System (PPMS) of Quantum design. The resistivity and Hall coefficient were also measured in this system.

**Measurements of SdH oscillations.**

The 2M WS$_2$ crystal was cleaved onto a silicon substrate. Au electrodes of standard Hall bar configuration were patterned by an electron-beam lithography process on isolated 2M WS$_2$ flakes. Thicknesses of the flakes were measured by AFM as 90 nm and 130 nm for samples #1 and #2, respectively.

A standard six-probe method was performed in a dilution refrigerator and superconducting magnet. Two crystals were measured with different geometry of electrical current. In-plane crystal axes of the samples were deduced from the anisotropy of the transport properties as well as the symmetry of the SdH frequencies as a function of rotation angle in a magnetic field.

Simulation of SdH frequencies were performed by using a program based on SKEAF code (*48*).

**First-principles calculations**

The first-principles calculations are carried out in the framework of the generalized gradient approximation (GGA) functional of the density functional theory through employing the Vienna ab initio simulation package (VASP) with projector augmented wave pseudopotentials (*49*). The experimental lattice constants are taken, and inner positions are obtained through full relaxation with a total energy tolerance $10^{-5}$ eV. The SOC effect is self-consistently included. Modified Becke-Johnson (mBJ) functional is employed for the double check and the electronic structure and topological nature remain unchanged.

**Explanation for the origin of the SdH frequency around 80 T**

We could not identify the origin of $F_†$ in **Figure 4e**. An almost two-dimensional angular dependence of frequency is confirmed in the rotation within $a^*$–$b$ plane, which is different from what is expected from β branch. Besides, the β oscillation is expected to be very weak due to the large $m_{band}/m_e$ = 1.54 and the large curvature, while $F_†$ has $m_{cyc}/m_e$ = 0.70(15). Thus, there seems to be no candidate frequency left in the bulk Fermi surfaces. It is possible that this frequency is from the topological surface states, although further investigation is necessary to draw a conclusion.

**Accession Codes**

CCDC 1853656 contains the supplementary crystallographic data for this paper. These data can be obtained free of charge via www.ccdc.cam.ac.uk/data_request/cif, or by emailing data_request@ccdc.cam.ac.uk, or by contacting The Cambridge Crystallographic Data Centre, 12 Union Road, Cambridge CB21EZ, UK; fax: +44 1223 336033.

## Supplementary materials

**Table S1.** The result of the EDS measurements for the $K_xWS_2(x\approx0.7)$
**Table S2.** Crystallographic Data for 2M $WS_2$.
**Table S3.** Crystal Data and Structural Refinement statistics for 2M $WS_2$.
**Table S4.** Anisotropic displacement parameters ($Å^2$) for 2M $WS_2$.
**Table S5.** Selected bond distances (Å) and bond angles (°) for 2M $WS_2$.
**Figure S1.** SEM image and EDS spectrum of the prepared 2M $WS_2$ crystals.
**Figure S2.** Heat measurement of bulk 2M $WS_2$
**Figure S3.** Temperature dependence and the anisotropy ratio of the upper critical magnetic field $H_{c2}$ of 2M $WS_2$ crystals.
**Figure S4.** The calculated phonon dispersion for 2M $WS_2$
**Figure S5.** Calculated GGA phonon density of state and electron-phonon spectral function $α^2F(ω)$ for 2M $WS_2$.
**Figure S6.** The density of states (DOS) and the projected DOS of 2M $WS_2$.

## References and Notes

1.  J. Kibsgaard, Z. B. Chen, B. N. Reinecke, T. F. Jaramillo, Engineering the surface structure of $MoS_2$ to preferentially expose active edge sites for electrocatalysis. *Nat. Mater.* **11**, 963-969 (2012).

2.  M. A. Lukowski *et al.*, Enhanced hydrogen evolution catalysis from chemically exfoliated metallic $MoS_2$ nanosheets. *J. Am. Chem. Soc.* **135**, 10274-10277 (2013).

3.  H. Li *et al.*, Activating and optimizing $MoS_2$ basal planes for hydrogen evolution through the formation of strained sulphur vacancies. *Nat. Mater.* **15**, 48-53 (2016).

4.  X. F. Qian, J. W. Liu, L. Fu, J. Li, Quantum spin Hall effect in two-dimensional transition metal dichalcogenides. *Science* **346**, 1344-1347 (2014).

5.  Z. Y. Fei *et al.*, Edge conduction in monolayer $WTe_2$. *Nat. Phys.* **13**, 677-682 (2017).

6.  L. Peng *et al.*, Observation of topological states residing at step edges of $WTe_2$. *Nat. Commun.* **8**, 659 (2017).

7.  S. Tang *et al.*, Quantum spin Hall state in monolayer 1T'-$WTe_2$. *Nat. Phys.* **13**, 683-687 (2017).

8.  P. Chen *et al.*, Large quantum-spin-Hall gap in single-layer 1T' $WSe_2$. *Nat. Commun.* **9**, 2003 (2018).

9. S. F. Wu *et al.*, Observation of the quantum spin Hall effect up to 100 kelvin in a monolayer crystal. *Science* **359**, 76-79 (2018).

10. A. A. Soluyanov *et al.*, Type-II Weyl semimetals. *Nature* **527**, 495-498 (2015).

11. Y. Sun, S. C. Wu, M. N. Ali, C. Felser, B. H. Yan, Prediction of Weyl semimetal in orthorhombic $MoTe_2$. *Phys. Rev. B* **92**, (2015).

12. J. Jiang *et al.*, Signature of type-II Weyl semimetal phase in $MoTe_2$. *Nat. Commun.* **8**, 13973 (2017).

13. K. Deng *et al.*, Experimental observation of topological Fermi arcs in type-II Weyl semimetal $MoTe_2$. *Nat. Phys.* **12**, 1105-1110 (2016).

14. L. Huang *et al.*, Spectroscopic evidence for a type II Weyl semimetallic state in $MoTe_2$. *Nat. Mater.* **15**, 1155-1160 (2016).

15. P. Li *et al.*, Evidence for topological type-II Weyl semimetal $WTe_2$. *Nat. Commun.* **8**, 13973 (2017).

16. I. Belopolski *et al.*, Discovery of a new type of topological Weyl fermion semimetal state in $Mo_xW_{1-x}Te_2$. *Nat. Commun.* **7**, 13643 (2016).

17. D. Xiao, G. B. Liu, W. X. Feng, X. D. Xu, W. Yao, Coupled Spin and Valley Physics in Monolayers of $MoS_2$ and Other Group-VI Dichalcogenides. *Phys. Rev. Lett.* **108**, (2012).

18. H. L. Zeng, J. F. Dai, W. Yao, D. Xiao, X. D. Cui, Valley polarization in $MoS_2$ monolayers by optical pumping. *Nat. Nanotech.* **7**, 490-493 (2012).

19. K. F. Mak, K. He, J. Shan, T. F. Heinz, Control of valley polarization in monolayer $MoS_2$ by optical helicity. *Nat. Nanotech.* **7**, 494-498 (2012).

20. X. D. Xu, W. Yao, D. Xiao, T. F. Heinz, Spin and pseudospins in layered transition metal dichalcogenides. *Nat. Phys.* **10**, 343-350 (2014).

21. H. T. Yuan *et al.*, Zeeman-type spin splitting controlled by an electric field. *Nat. Phys.* **9**, 563-569 (2013).

22. D. MacNeill *et al.*, Breaking of Valley Degeneracy by Magnetic Field in Monolayer $MoSe_2$. *Phys. Rev. Lett.* **114**, (2015).

23. B. Sipos *et al.*, From Mott state to superconductivity in $1T-TaS_2$. *Nat. Mater.* **7**, 960-965 (2008).

24. J. T. Ye *et al.*, Superconducting Dome in a Gate-Tuned Band Insulator. *Science* **338**, 1193-1196 (2012).

25. T. Yokoya *et al.*, Fermi surface sheet-dependent superconductivity in $2H-NbSe_2$. *Science* **294**, 2518-2520 (2001).

26. A. H. Castro Neto, Charge density wave, superconductivity, and anomalous metallic behavior in 2D transition metal dichalcogenides. *Phys. Rev. Lett.* **86**, 4382-4385 (2001).

27. Y. J. Yu *et al.*, Gate-tunable phase transitions in thin flakes of 1T-TaS$_2$. *Nat. Nanotech.* **10**, 270-276 (2015).

28. D. Costanzo, S. Jo, H. Berger, A. F. Morpurgo, Gate-induced superconductivity in atomically thin MoS$_2$ crystals. *Nat. Nanotech.* **11**, 339-344 (2016).

29. X. X. Xi *et al.*, Ising pairing in superconducting NbSe$_2$ atomic layers. *Nat. Phys.* **12**, 139-143 (2016).

30. J. M. Lu *et al.*, Evidence for two-dimensional Ising superconductivity in gated MoS$_2$. *Science* **350**, 1353-1357 (2015).

31. M. H. Whangbo, E. Canadell, Analogies between the Concepts of Molecular Chemistry and Solid-State Physics Concerning Structural Instabilities - Electronic Origin of the Structural Modulations in Layered Transition-Metal Dichalcogenides. *J. Am. Chem. Soc.* **114**, 9587-9600 (1992).

32. E. Benavente, M. A. Santa Ana, F. Mendizabal, G. Gonzalez, Intercalation chemistry of molybdenum disulfide. *Coord. Chem. Rev.* **224**, 87-109 (2002).

33. M. N. Ali *et al.*, Large, non-saturating magnetoresistance in WTe$_2$. *Nature* **514**, 205-208 (2014).

34. Y. Qi *et al.*, Superconductivity in Weyl semimetal candidate MoTe$_2$. *Nat Commun* **7**, 11038 (2016).

35. D. F. Kang *et al.*, Superconductivity emerging from a suppressed large magnetoresistant state in tungsten ditelluride. *Nat. Commun.* **6**, 7804 (2015).

36. F. Wypych, K. Sollmann, R. Schollhorn, Metastable Layered Chalcogenides 1T-MoS$_2$, 2M-WS$_2$ and 1T-Mo$_{1/2}$W$_{1/2}$S$_2$ - Electrochemical Study on Their Intercalation Reactions. *Materials Research Bulletin* **27**, 545-553 (1992).

37. W. G. Dawson, D. W. Bullett, Electronic-Structure and Crystallography of MoTe$_2$ and WTe$_2$. *Journal of Physics C-Solid State Physics* **20**, 6159-6174 (1987).

38. Q. Liu *et al.*, Stable Metallic 1T-WS$_2$ Nanoribbons Intercalated with Ammonia Ions: The Correlation between Structure and Electrical/Optical Properties. *Adv. Mater.* **27**, 4837-4844 (2015).

39. E. Morosan, L. Li, N. P. Ong, R. J. Cava, Anisotropic properties of the layered superconductor Cu$_{0.07}$TiSe$_2$. *Phys. Rev. B* **75**, (2007).

40. C. Shekhar *et al.*, Extremely large magnetoresistance and ultrahigh mobility in the topological Weyl semimetal candidate NbP. *Nat. Phys.* **11**, 645-649 (2015).

41. Z. K. Liu *et al.*, A stable three-dimensional topological Dirac semimetal Cd$_3$As$_2$. *Nat. Mater.* **13**, 677-

681 (2014).

42. A. Molina-Sanchez, L. Wirtz, Phonons in single-layer and few-layer MoS$_2$ and WS$_2$. *Phys. Rev. B* **84**, 155413 (2011).

43. A. Subedi, L. J. Zhang, D. J. Singh, M. H. Du, Density functional study of FeS, FeSe, and FeTe: Electronic structure, magnetism, phonons, and superconductivity. *Phys. Rev. B* **78**, (2008).

44. N. Marzari, D. Vanderbilt, Maximally localized generalized Wannier functions for composite energy bands. *Phys. Rev. B* **56**, 12847-12865 (1997).

45. H. J. Zhang *et al.*, Electronic structures and surface states of the topological insulator Bi$_{1-x}$Sb$_x$. *Phys. Rev. B* **80**, (2009).

46. L. Fu, C. L. Kane, Superconducting proximity effect and Majorana fermions at the surface of a topological insulator. *Phys. Rev. Lett.* **100**, (2008).

47. S. Sasaki *et al.*, Topological Superconductivity in Cu$_x$Bi$_2$Se$_3$. *Phys. Rev. Lett.* **107**, (2011).

48. P. M. C. Rourke, S. R. Julian, Numerical extraction of de Haas-van Alphen frequencies from calculated band energies. *Comput. Phys. Commun.* **183**, 324-332 (2012).

49. G. Kresse, D. Joubert, From ultrasoft pseudopotentials to the projector augmented-wave method. *Phys. Rev. B* **59**, 1758-1775 (1999).

# Acknowledgements

# Funding

This work was financially supported by National key R&D Program of China (Grant 2016YFB0901600), Science and Technology Commission of Shanghai (Grant 16JC1401700), National Science Foundation of China (Grant 51672301), the Key Research Program of Chinese Academy of Sciences (Grants No. QYZDJ-SSW-JSC013 and KGZD-EW-T06), Innovation Project of Shanghai Institute of Ceramics (Grant No. Y73ZC6160G), the National Natural Science Foundation of China (No. 11674165), the Fok Ying-Tong Education Foundation of China (Grant No. 161006), the Youth Innovation Promotion Association of the Chinese Academy of Sciences (Grants No. 2015187), and CAS Center for Excellence in Superconducting Electronics, partially supported by JSPS KAKENHI Grant No. JP17H07349.

# Author contributions

Y. Q. Fang and J. Pan synthesized the 2M WS$_2$ compound. D. Q. Zhang and H. J. Zhang calculated

the electronic structure and topological state. D. Wang carried out the single-crystal refinement. Y. Q. Fang, J. Pan, Y. H. Ma and G. Mu and W. Zhao performed the physical transport measurements including magnetic, electrical and heat capacity measurements. Hishiro T. Hirose, Taichi Terashima and Shinya Uji conducted the measurements of SdH oscillations. Y. H. Yuan, W. Li, Q. K. Xue, Z. Tian and J. M. Xue carried out the STM measurements and analyzed the data. Y. Q. Fang, J. Pan, D. Q. Zhang, H. J. Zhang, G. Mu and F. Q. Huang analyzed the data and wrote the paper. F. Q. Huang conceived and coordinated the project, and is responsible for the infrastructure and project direction. All authors discussed the results and commented on the manuscript.

## Additional information

## Competing financial interests

The authors declare no competing financial interests.

# Figures and Tables

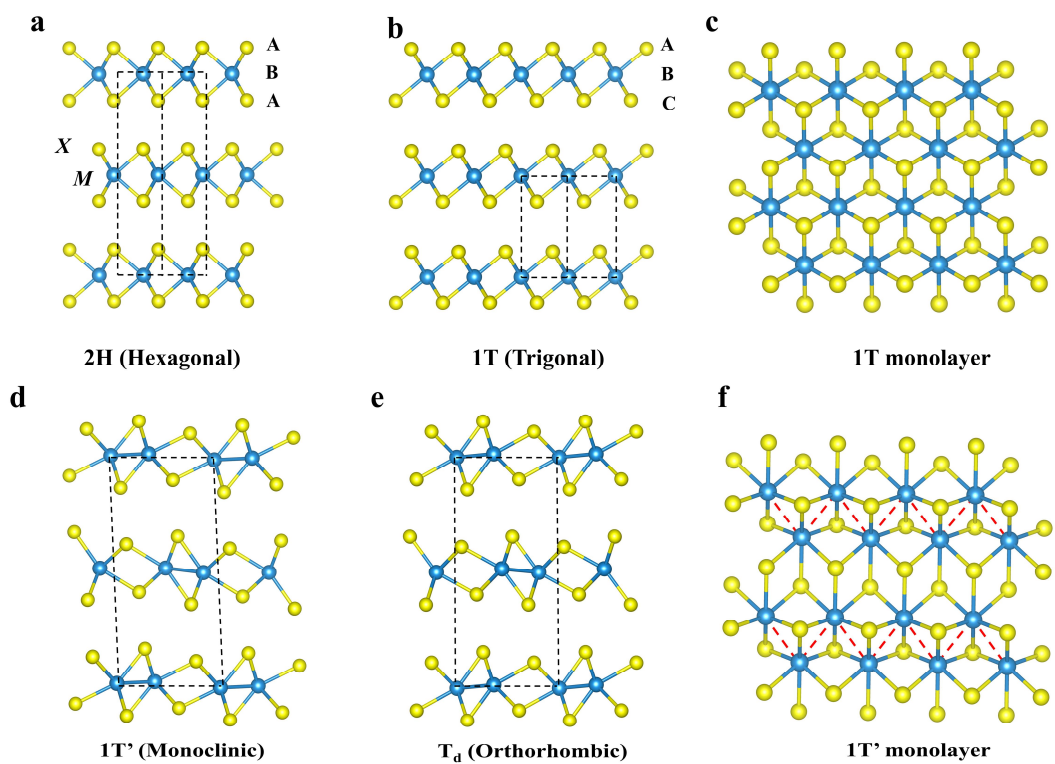

**Figure 1. Typical crystal structures of TMDs. a-f**, 2H phase (**a**), 1T phase (**b**), 1T monolayer (**c**), 1T' phase (**d**), $T_d$ phase (**e**) and 1T' monolayer (**f**). The blue and yellow spheres represent *M* (Mo or W) and *X* (S, Se or Te) atoms, respectively.

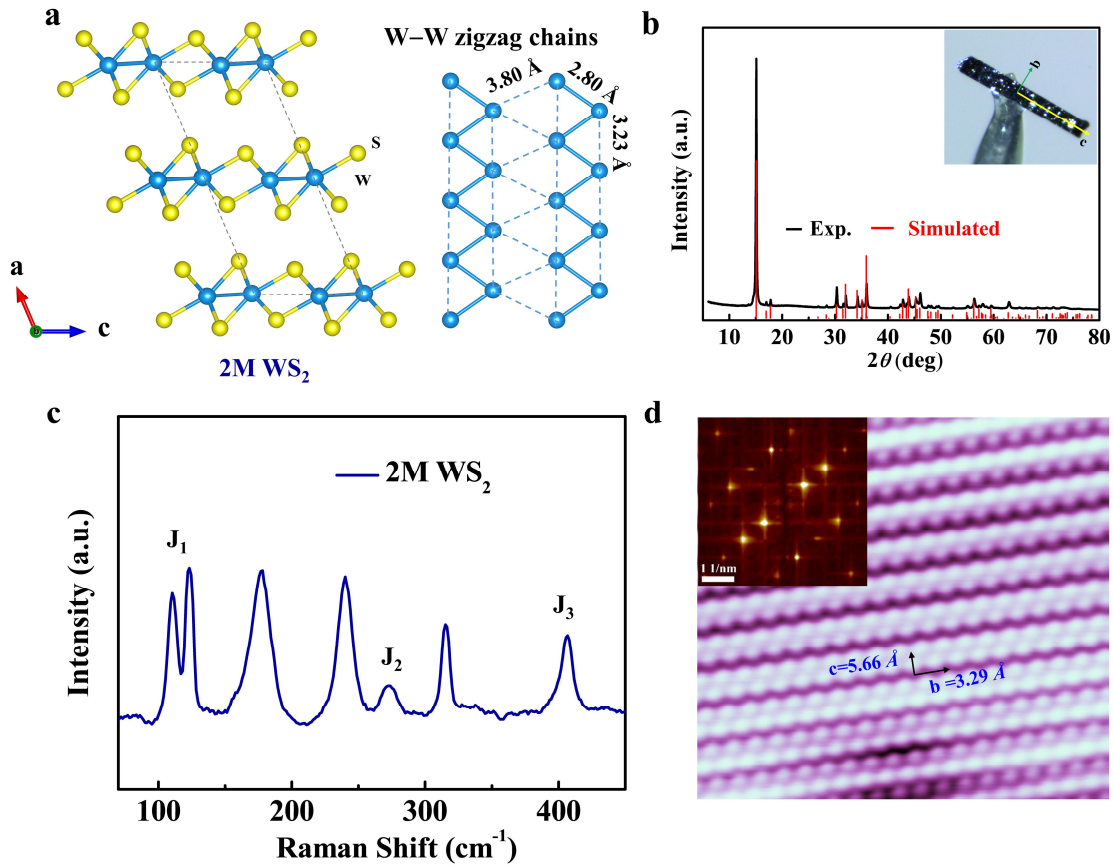

**Figure 2. Crystal structure characterizations of 2M WS$_2$. a,** crystal structure of 2M WS$_2$. **b,** The powder X-ray diffraction pattern of 2M WS$_2$ crystals, where the red lines were simulated from single crystal structure of 2M WS$_2$. Inset: the optical image of as-grown 2M WS$_2$ crystal. **c,** The Raman spectrum of the bulk 2M WS$_2$ crystal. **d,** The STM topography of 2M WS$_2$ crystal. Inset: the corresponding FFT pattern.

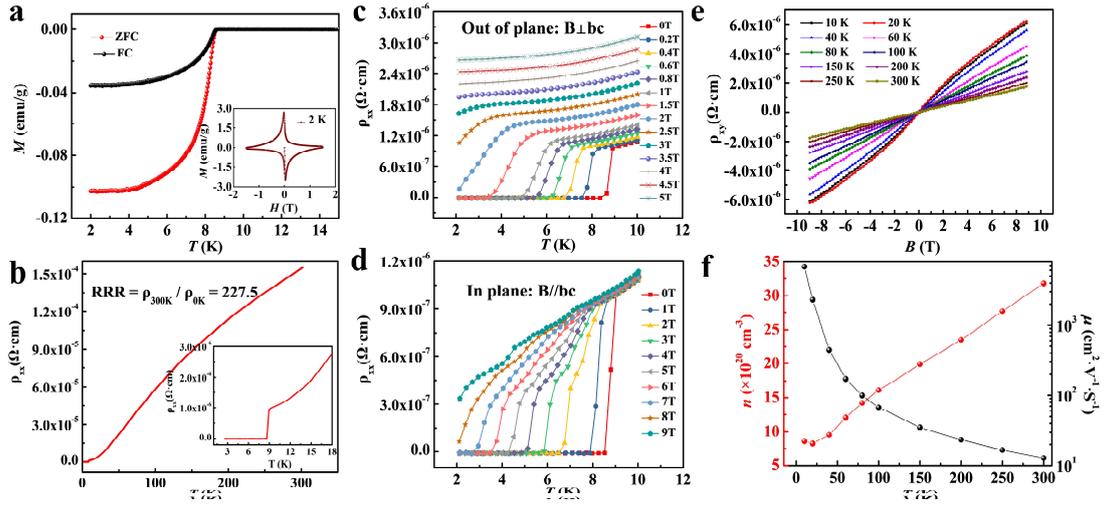

**Figure 3. Superconductivity of 2M WS$_2$. a,** Temperature dependence of magnetic susceptibility for 2M WS$_2$ under the magnetic field of 5 Oe with zero-field-cooling curve (red) and field-cooling curve (black). The inset exhibits the magnetic hysteresis of the sample measured at 2K. **b,** Temperature dependence of resistivity of 2M WS$_2$. **c,** The temperature dependence of resistivity under different out-of-plane magnetic fields. **d,** The temperature dependence of resistivity under different in-plane magnetic fields. **e,** The magnetic field dependence of Hall resistivity under different temperature. **f,** The temperature dependence of carrier concentration and carrier mobility, respectively.

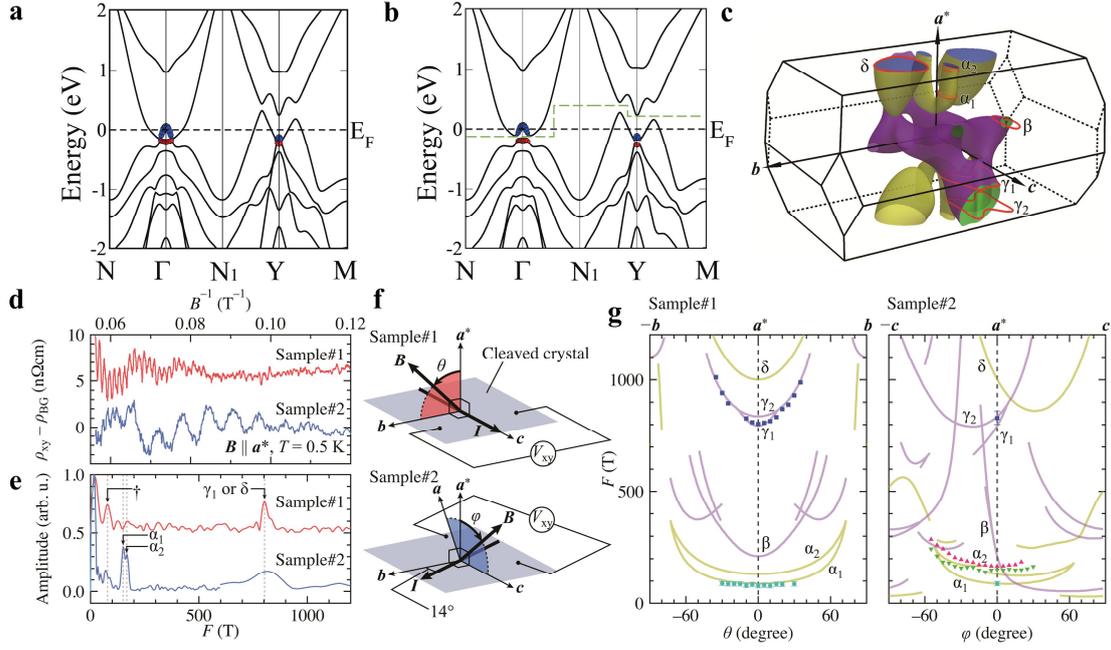

**Figure 4. Electronic structure of 2M WS$_2$. a** and **b,** Band structure without and with spin-orbit coupling. The blue closed circles present the projection of *p* orbital of S atom, and the red closed circles present the projection of *d* orbital of W atom, indicating the band inversion at the Γ point. The green dashed line in **b** is schematic curving Fermi level. **c,** The 3-dimensional Fermi surfaces of bulk states indicate the coexisting of hole and electron pockets. **d,** Magnetic field dependence of SdH oscillations in $\rho_{xy}$ as a function of inverse field $B^{-1}$ for samples #1 and #2 having different configuration of electrical current. **e,** Fourier transforms of the SdH oscillations. For clarity, spectra are vertically shifted and different field ranges are used: 8–17.8 T for sample #1, 6–17.8 T ($F$ < 600 T) and 14–17.8 T ($F$ > 600 T) for sample #2. **f,** Schematic diagrams of the measurement configuration in samples #1 and #2. In sample #2, because the electrical current is tilted with respect to the major axes, $\rho_{xy}$ effectively contains both components of Hall resistivity and magnetoresistivity. **g,** Angular dependence of the SdH frequencies of both experiments (solid marks) and simulations (solid lines). The simulated frequencies of the hole (electron) Fermi surface is drawn in yellow (purple).

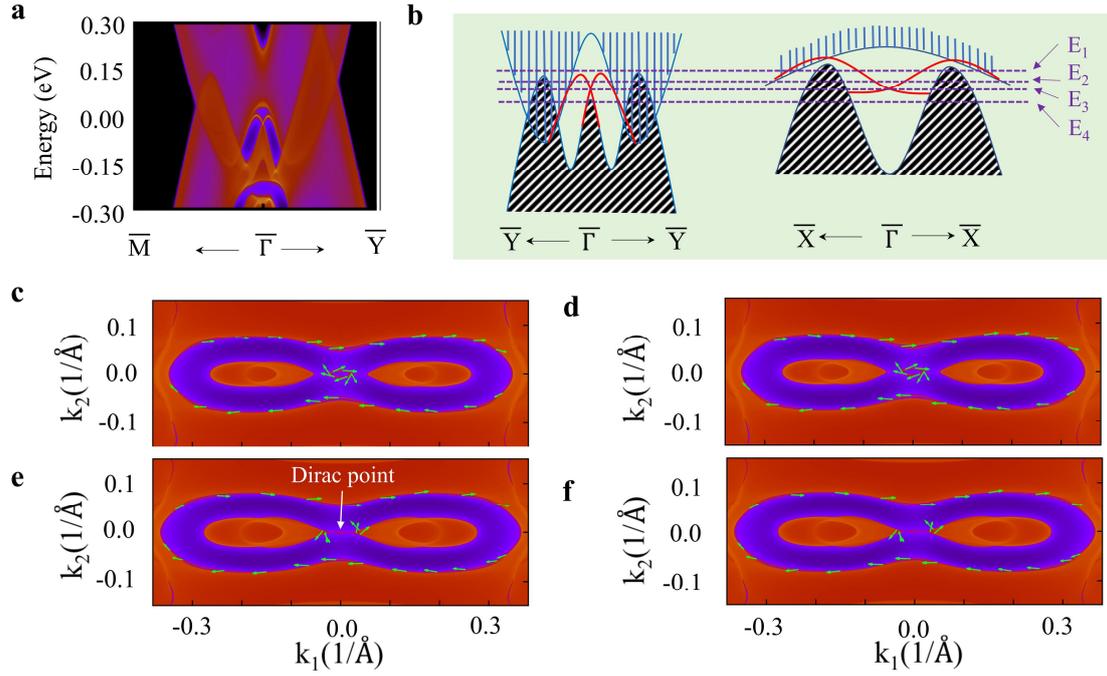

**Figure 5. Topological Surface states of 2M WS$_2$. a**, The bulk and surface states are projected to the (100) surface. The bulk states are consistent with the semimetal picture. **b**, The schematic of band structures on the (100) surface. The red lines are the schematic surface states. The four dashed lines indicate different energy level E1, E2, E3 and E4. **c-f**, The Fermi surface at the energy level E1 in **c**, at the energy level E2 in **d**, at the energy level E3 in **e**, and at the energy level E4 in **f**. The green arrows are the spin texture.

**Table 1.** The parity products of valence bands defined by the curving Fermi level at each time-reversal-invariant point.

| Γ (000) | N (π00) | $N_1$ (0π0) | Z (00π) |
|---|---|---|---|
| − | + | + | + |
| Y (ππ0) | M (π0π) | $Y_1$ (0ππ) | L (πππ) |
| + | + | + | + |
| Topological invariant $Z_2$ (1;000) | | | |